\documentclass[letterpaper]{article} 
\usepackage{aaai24}  
\usepackage{times}  
\usepackage{helvet}  
\usepackage{courier}  
\usepackage[hyphens]{url}  
\usepackage{graphicx} 
\usepackage{booktabs}
\usepackage{amsmath}
\usepackage{amssymb}

\usepackage{natbib}  
\usepackage{caption} 
\usepackage{algorithm}
\usepackage{algorithmic}
\usepackage{comment}
\usepackage{latexsym}
\usepackage{enumitem}
\usepackage{tabularx}
\usepackage{booktabs}
\usepackage{multicol}
\usepackage{multirow}
\usepackage{pgfplots}
\usepackage{todonotes}
\usepackage{svg}
\pgfplotsset{width=5cm,compat=1.8}
\setlist[enumerate]{itemsep=0mm}
\usepackage{newfloat}
\usepackage{listings}
\DeclareCaptionStyle{ruled}{labelfont=normalfont,labelsep=colon,strut=off} 
\floatstyle{ruled}
\newfloat{listing}{tb}{lst}{}
\floatname{listing}{Listing}
\title{$\mathcal{K}$-PERM: Personalized Response Generation Using Dynamic Knowledge Retrieval and Persona-Adaptive Queries}
\author{
    Kanak Raj\equalcontrib\textsuperscript{\rm 1},
    Kaushik Roy\equalcontrib\textsuperscript{\rm 2},
    Vamshi Bonagiri\textsuperscript{\rm 3},
     Priyanshul Govil\textsuperscript{\rm 3}, 
    Krishnaprasad Thirunarayanan\textsuperscript{\rm 4}, 
    Manas Gaur\textsuperscript{\rm 3}
}
\affiliations{
    \textsuperscript{\rm 1}HealthCareNLP Softech LLP\\
    \textsuperscript{\rm 2}Artificial Intelligence Institute, University of South Carolina\\
    \textsuperscript{\rm 3}University of Maryland, Baltimore County\\
    \textsuperscript{\rm 4}Wright State University

}

\begin{document}

\maketitle

\begin{abstract}
Personalizing conversational agents can enhance the quality of conversations and increase user engagement. However, they often lack external knowledge to appropriately tend to a user's persona. This is particularly crucial for practical applications like mental health support, nutrition planning, culturally sensitive conversations, or reducing toxic behavior in conversational agents. To enhance the relevance and comprehensiveness of personalized responses, we propose using a two-step approach that involves (1) selectively integrating user personas and (2) contextualizing the response with supplementing information from a background knowledge source. We develop \textbf{$\mathcal{K}$-PERM} (\textbf{K}nowledge-guided \textbf{PE}rsonalization with \textbf{R}eward \textbf{M}odulation), a dynamic conversational agent that combines these elements. $\mathcal{K}$-PERM achieves state-of-the-art performance on the popular FoCus dataset, containing real-world personalized conversations concerning global landmarks. We show that using responses from $\mathcal{K}$-PERM can improve performance in state-of-the-art LLMs (GPT 3.5) by 10.5\%, highlighting the impact of $\mathcal{K}$-PERM for personalizing chatbots. \footnote{Our code is released to the public for further explorations: \texttt{\small https://github.com/kanak8278/DialogKPERM}}
\end{abstract}

\section{Introduction}

Recent trends in Large Language Models (LLMs) have demonstrated remarkable abilities in conversational AI \cite{jo2023understanding}\cite{shin2023planfitting}. However, personalization is a potential area for LLMs that requires improvement \cite{zhang2023memory}. Personalization in conversational AI can go beyond chit-chat conversations and aid in user engagement by understanding their personas better and providing accurate responses \cite{bender2021dangers}\cite{joshi2017personalization}. 

Prior research on personalization has primarily focused on casual conversations, emphasizing details such as a user's preferences. 
The lack of external knowledge hinders a model's ability to adapt to different personas \cite{deshpande2023toxicity}. Therefore, recent shifts in chatbot personalization utilize both persona information and knowledge \cite{qian2021learning} \cite{liu2023recap}. However, identifying a suitable context aligned with user preferences during a conversation remains a significant challenge for current LLMs. 

While using various prompting methods may allow a user to steer LLMs toward desired behavior, they only work at an utterance level. This may not be feasible for longer conversations, as the context often shifts \cite{shuster2021retrieval}. Therefore, we require the chatbot to learn how to retrieve appropriate content based on a user's query and assess whether a response requires contextualization (retrieval of meta-information) and personalization (selecting appropriate persona, if necessary). 

To address this issue, we propose using \textbf{K}nowledge-guided \textbf{PE}rsonalization of response generation with \textbf{R}eward \textbf{M}odulation (\textbf{$\mathcal{K}$-PERM}). \textbf{$\mathcal{K}$-PERM} uses dynamic knowledge retrieval along with personalization to improve a machine's adaptive capabilities to different personas and contexts. Our Personalized Response Generation task involves two major components: 
\begin{enumerate}[itemsep=0mm]
    \item \textbf{Understanding Conversation Context:} We use Dense Passage Retrieval (DPR) \cite{karpukhin2020dense} to select the most pertinent information from a larger text corpus containing real-world information. 
    \item \textbf{Incorporating Appropriate Personas:} We introduce a selector module capable of choosing a persona that aligns with the user query. We model persona selection as a multiple-choice question-answering task, which includes an option to opt for ``no-persona'' in generic cases. 
\end{enumerate}

\begin{figure*}[t]
  \centering
  \includegraphics[width=0.7\textwidth, trim=5cm 6cm 5cm 2cm]{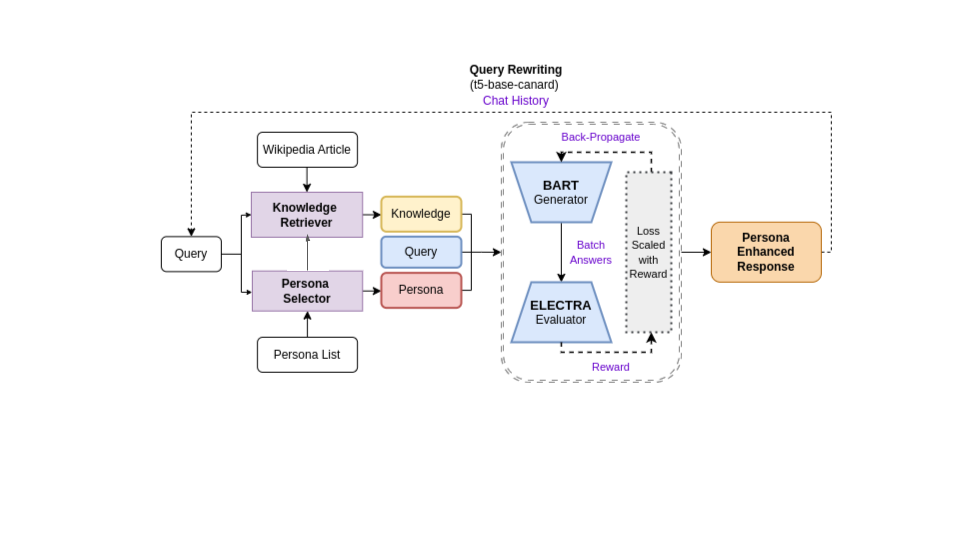}
  \caption{\footnotesize $\mathcal{K}$-PERM Model Architecture Overview. The model architecture comprises a Persona Selector and Knowledge Extractor, which leverages the history and question prompt to identify pertinent persona and knowledge.} 
  \label{fig:model_arch}
\end{figure*}

To the best of our knowledge, our study is the first to dynamically and flexibly incorporate knowledge, persona, and conversation history into a learning strategy that can be used to train or guide LLMs to be personalized. Our framework is model-agnostic and can be used to train conversational agents for personalized response generation in open-domain settings. $\mathcal{K}$-PERM outperforms prior attempts in personalization despite being 24 times smaller model. We also augment GPT 3.5 with $\mathcal{K}$-PERM's responses and show a huge improvement of 10.5\% when compared to GPT3.5 in a zero-shot setting. Our results show that knowledge-guided learning can help guide conversational agents toward better personalization. 


\section{Methodology}

Let $U^t = \{u^t_1,u^t_2,\hdots,u^t_H\}$ be the history of $H$ user utterances on a topic $t$. Each utterance $u^t_i \in U^t$ is a question and response pair denoted as $(q^t_i,r^t_i)$. The $q^t_i$ are questions by the user, specific to the topic $t$, and the $r^t_i$ are responses that are either tailored according to a set of $n$ user personas $P = \{p_1,p_2,\hdots,p_n\}$ or generic (no-persona). Our goal is to model the probability of the latest user response $r^t_H$ given a set of $K$ passages, denoted by $\mathcal{Z}^{U^t}_K$, relevant to the utterance history $U^t$ (the passages are drawn from an external knowledge source, e.g., a document store). Thus, our goal is to learn the probability distribution 
\begin{equation}\label{eqn:prob_model}
 \mathbb{P}_\theta(r^t_H \mid \mathcal{Z}^{U^t}_K)   
\end{equation}
where $\theta$ are the parameters of the probability distribution.

$\mathbb{P}_\theta$ can be any auto-regressive language model capable of generating sentences token-wise. Since the user is likely to have responded according to their set of personas, we train a persona selector module $P_\text{select}$, that takes as input the user's utterance history $U^t$, and the set of passages $\mathcal{Z}^{U^t}_K$, and outputs one or more personas from $P$ (denoted as $P'$) for customizing responses using $\mathbb{P}_\theta$.  Therefore, Equation \eqref{eqn:prob_model} is modified as
\begin{align}\label{eqn:prob_model2}
\small
\begin{split}
    P' =\ & P_\text{select}(\mathcal{Z}^{U^t}_K,P), \quad P' \subseteq P,\\
    &\mathbb{P}_\theta(r^t_H \mid \mathcal{Z}^{U^t}_K, P') 
\end{split}
\end{align}


\section{$\mathcal{K}$-PERM}
Figure \ref{fig:model_arch} explains the entire model architecture of $\mathcal{K}$-PERM. Utilizing the conversation history $U^t$, we access a document store, retrieve pertinent passages, and rank them based on their relevance. This allows us to leverage the retrieved information to select compatible user personas and generate personalized responses. Our method personalizes based on the personas in $P'$. If $P' = \emptyset$, generic responses are generated. 

\subsection{Knowledge Retriever} For dynamically retrieving passages based on $U^t$, we built upon a process called DPR -- $P_\text{select}$ in Equation \eqref{eqn:prob_model2}. DPR uses semantic similarity search to retrieve passages from a vectorized database. This allows us to go beyond an \textit{exact-match} by retrieving passages that can answer questions by automatically adapting to the input queries in $U^t$ -- \textit{what is?} \textit{what if?} \textit{what could be?}

We improve DPR in two ways. First, we utilize a Sentence-BERT model for performing a \textbf{retrieve-rank process using a paired cross-encoder and bi-encoder}. A cross-encoder retrieves a set of passages given the last query $q^t_H \in U^t$, and subsequently, the bi-encoder ranks and selects the top-$K$ passages to result in $\mathcal{Z}^{U^t}_K$. We create dense encodings of the passages in $z_j \in \mathcal{Z}^{U^t}_K$ using the MPNET model from the Sentence-BERT transformer family \cite{reimers2019sentencebert}. Likewise, the encoding for $q^t_H$ is represented as $z_H$. We fine-tune Mpnet on our dataset before using it to obtain dense encodings (refer Appendix{mpnet-ft} {D}) \cite{song2020mpnet}.

\subsection{Persona Selector ($P_\text{select}$) \label{subsubsec:pselect}}
We model persona selection as a commonsense inference task, conditioned upon the query knowledge $\mathcal{Z}^{U^t}_K$ retrieved using the information in $q^t_H$ and the set of user-personas $P$, formally written as $P' = P_\text{select}(\mathcal{Z}^{U^t}_K, P)$ as shown in Equation \eqref{eqn:prob_model2}. The dataset contains the ground truth for the user's personas corresponding to the responses in the utterance history $U^t$. Using this, we train the $P_\text{select}$ model as a multi-label classifier and sample the top-2 classes from the resulting logits ($|P'| = 2$). For the base $P_\text{select}$ model, we empirically observed XLNET to give the highest performance \cite{yang2019xlnet}.

\subsection{Response Generation through Reward Modulation} The response is generated by pairing a BART(Base) generator with an ELECTRA(Base) evaluator that measures the similarity between the generated response and the ground truth \cite{clark2019electra}. We introduce a balancing reward function $(R_i)$ modulating generative capabilities (e.g., coherence) of the BART model and high fidelity to the ground truth responses (in terms of matched words).

\subsection{Reward Function} Consider the ground truth response for a query $q^t_i$ to be $r^t_i$, which consists of $n$ tokens, where each token is indexed using $t_i$. Let $r^t_k$ be the $k^{\text{th}}$ response in the generated response list comprising $m$ tokens, where each token is indexed using $t_j$. We generate BERT encodings for each word in the response vectors, both for the $n$ words in $r^t_i$ and the $m$ words in $r^t_k$. The reward ($R_i$) is

\begin{align}
\begin{split}\label{eqn:reward_eqn}
{R_i}=\ & \alpha \cdot \operatorname{BLEU}\left(r^t_i, r^t_k\right) +\\
&(1-\alpha)\sum_{(t_i \in r^t_i,t_j \in r^k_i)} \max_{t_i} \operatorname{WMD}\left(t_i, t_j\right)
\end{split}
\end{align}

where $\text{WMD}$ denotes the Word Mover Distance \cite{kusner2015word} and $\alpha[*] \in [0,1]$ balances between a well-generated response by BART (given by the WMD distance) and closeness to the ground truth responses (given by the BLEU score).

\subsection{Persona-tailored Reward Function } In addition to producing syntactically sound responses and responses that are close to the ground truth, the responses need to be tailored to user-personas, i.e., the output of the $P_\text{select}$ model. Thus, we modify Equation \eqref{eqn:reward_eqn} as follows:

\begin{align}\label{eqn:reward_eqn2}
\begin{split}
{R_i} =\ & \alpha \operatorname{BLEU}\left(r^t_i, r^t_k\right)+\\
&\beta\sum_{(t_i \in r^t_i,t_j \in r^k_i)} \max_{t_i} \operatorname{WMD}\left(t_i, t_j\right) +\\
&\gamma \cdot \text{Loss}(P_\text{select},P_\text{gt}),\quad \alpha+\beta+\gamma = 1
\end{split}
\end{align}
where $gt$ is the ground truth, and $\text{Loss}(P_\text{select},P_\text{gt})$ refers to the loss function, i.e., the training error in the persona selected and the ground truth persona (if present).

\section{Experiments}

We utilize the FoCus dataset developed by \citet{jang2022call} for our experiments, as it contains customized answers built with persona and Wikipedia knowledge instead of just persona (\cite{zhang2018personalizing}). See Appendix \ref{dataset} for dataset details. Our experiments use BART as the language model ($\mathbb{P}_\theta$). Fine-tuning details are described in Appendix C.

\subsection{Evaluation Criteria} We used Rouge--1/2/L/L-Sum and BLEU scores to evaluate $\mathcal{K}$-PERM. In addition, we use two transformer-based metrics for evaluating natural language generation: BERTScore (BF1: BERTScore F1-score) and NUBIA, which measure semantic relations, contractions, irrelevancy, and logical agreement \cite{zhang2020bertscore,kane2020nubia}. Semantic relations evaluate whether the generated text is relevant and coherent and maintains the intended meaning and context of the input query. Measuring these characteristics for our proposed model is important to ensure LLMs' correct merging of knowledge and persona.

\subsection{Baselines} We compare our model with two baselines -- \textbf{(1) GODEL}: A large pre-trained Transformer-based encoder-decoder model for goal-directed dialogues similar to FoCus \cite{peng2022godel}. We used the pre-trained GODEL model and enhanced it with personalization by incorporating a persona selected by our persona selector model; \textbf{(2)} $\mathbf{BART}_\textbf{FoCus}$: We utilized the BART model provided with the FoCus dataset \cite{jang2022call}. We fine-tuned this model ($\text{BART}_\text{FoCus}$) using our training and validation sets for a fair comparison. Results are reported on our test set, although it was not made available by the authors of the FoCus dataset.

\subsection{Results and Discussion}
Table \ref{tab:comparison} compares three models: $\text{BART}_\text{FoCus}$ (406 Million Parameters), GODEL (6 Billion parameters), and $\mathcal{K}$-PERM (250 Million parameters). The results show that the pre-trained version of GODEL, without personalization, performed worse than $\text{BART}_\text{FoCus}$. However, when GODEL incorporated our persona selector model, it achieved a syntactic similarity closer to $\text{BART}_\text{FoCus}$, suggesting that the persona-based approach improved GODEL's syntactic quality. In terms of semantic similarity measured by BERTScore, GODEL with persona outperformed $\text{BART}_\text{FoCus}$, indicating that GODEL, when utilizing personas, generated responses that were more semantically similar to the desired outputs. $\mathcal{K}$-PERM significantly outperformed GODEL in terms of syntactic generation quality and semantic similarity. Ablation studies (Appendix table \ref{tab:ablation}) on $\mathcal{K}$-PERM showed that using the ground persona and our persona selector resulted in the highest quality generation both syntactically and semantically. However, there were cases where $\mathcal{K}$-PERM ignored the persona for specific queries requiring personalization, resulting in errors compared to using the ground persona. Despite this limitation, the knowledge retriever used in $\mathcal{K}$-PERM demonstrated competitive performance, relying on information from Wikipedia articles rather than handcrafted knowledge in the FoCus dataset.

Another set of experiments with $\mathcal{K}$-PERM using NUBIA as the metric (see Table \ref{tab:ablation}) showed that using personas yielded better semantic relations, logical agreement, lower contradiction, and lower irrelevancy compared to providing all personas simultaneously. When using retrieved knowledge, $\mathcal{K}$-PERM achieved a higher NUBIA score compared to using handcrafted knowledge in $\text{BART}_\text{FoCus}$. In a blind assessment of responses obtained from 5 different systems, generated across 90 user queries with varying numbers of personas, $\mathcal{K}$-PERM consistently outperformed the competition. It achieved the top position in 54 query cases, showcasing its remarkable performance in contrast to GODEL and GPT 3.5, as depicted in Appendix Figure \ref{fig:qual_eval}.

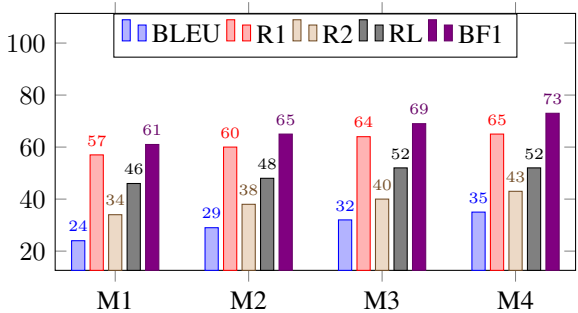
\begin{figure}
    \centering
    \begin{tikzpicture} 
\begin{axis}  
[  
    ybar, ymax=100,
    enlargelimits=0.15, 
    bar width=5pt,
    width=8.5cm, height=5cm,
    legend style={at={(0.5,1.0)},
            anchor=north,legend columns=-1},     
    ylabel={}, 
    symbolic x coords={M1, M2, M3, M4},  
    xtick=data,
    nodes near coords,
    every node near coord/.append style={font=\tiny, inner xsep=1.5pt},
   nodes near coords align={vertical}  
    ]  
\addplot coordinates {(M1, 24.0) (M2, 29) (M3, 32) (M4, 35)};  
\addplot coordinates {(M1, 57) (M2, 60) (M3, 64) (M4, 65)};  
\addplot coordinates {(M1, 34) (M2, 38) (M3, 40.0) (M4, 43)};
\addplot coordinates {(M1, 46) (M2, 48) (M3, 52.0) (M4, 52.0)};  
\addplot coordinates {(M1, 61) (M2, 65) (M3, 69) (M4, 73)};  
\legend{BLEU, R1, R2, RL, BF1}  
  
\end{axis}  
\end{tikzpicture} 
    \caption{\footnotesize $\mathcal{K}$-PERM improves personalization  in GPT 3.5 via zero-shot prompting. This experiment aimed to assess the performance improvement of GPT 3.5 when combined with $\mathcal{K}$-PERM. M1 is GPT 3.5 and M2, M3, and M4 represents zero-shot prompting of GPT 3.5 using responses from $\mathcal{K}$-PERM with (All P+$Z_k$), (GP+$Z_k$), and ($P_\text{select}$+$Z_k$) respectively.}
    \label{fig:augmentation}
\end{figure}

Finally, we use $\mathcal{K}$-PERM to augment a state-of-the-art LLM, GPT3.5. Our results in Figure \ref{fig:augmentation} show that augmenting GPT3.5 with $\mathcal{K}$-PERM improves its performance significantly (10.5\%), highlighting the advantage of $\mathcal{K}$-PERM in personalization.

\begin{table}[t]
\tiny
  \centering
  \begin{tabular}{p{2.5cm}ccccccc}
    \toprule
    Models & BLEU & R1 & R2 & RL  & BF1 \\
    \midrule
    ptGODEL & 5.21 & 25.82 & 9.87 & 21.2  & 43.86 \\
    ptGODEL ($P_\text{select}$) & 6.18 & 30.02 & 13.77 & 26.11  & 44.34 \\
    $\text{BART}_\text{FoCus}$ & 6.24 & 30.98 & 14.22 & 26.81  & 43.85 \\ \midrule[1pt]
    $\mathcal{K}$-PERM (without P, GK) & 2.14 & 24.68 & 9.46 & 21.89  & 44.15 \\
    $\mathcal{K}$-PERM (GK only) & 11.2 & 31.14 & 15.49 & 26.73  & 43.26 \\
    $\mathcal{K}$-PERM (GP only) & 2.4 & 27.47 & 12.03 & 22.37  & 44.35 \\ 
    \midrule[1pt]
    $\mathcal{K}$-PERM (All P+$\mathcal{Z}_k$) & 11.22 & 35.19 & 19.27 & 31.18 & 43.45 \\
    $\mathcal{K}$-PERM {(GP+$\mathcal{Z}_k$)} & \textbf{14.72} & \textbf{43.09} & \textbf{25.43} & \textbf{37.95}  & \textbf{47.36} \\
    $\mathcal{K}$-PERM ($P_\text{select}$+$\mathcal{Z}_k$) & \underline{12.01} & \underline{37.53} & \underline{23.96} & \underline{33.47}  & \underline{46.06} \\
    
    \bottomrule
  \end{tabular}
\caption{\footnotesize Performance of $\mathcal{K}$-PERM on FoCUS dataset. SP:$P_{select}$, GP: Ground Persona, $\mathcal{Z}_k$: Retrieved Knowledge, P: Persona, GK: Ground Truth Knowledge. Bold-face: Best and Underlined is 2nd best. pt: pre-trained.}
\label{tab:comparison}
\end{table}

\footnotetext{Score were rounded off in Figure 2 for visibility.}
\section{Conclusion}
We developed $\mathcal{K}$-PERM, a practical and comprehensive method for personalized response generation, showcasing its superior performance compared to the baselines. Specifically, $\mathcal{K}$-PERM responses align with human-curated responses on the FoCus dataset. Despite being a simpler language model compared to ChatGPT, $\mathcal{K}$-PERM ranked second in matching generated responses to the ground truth in FoCus. The reward modulation setting in $\mathcal{K}$-PERM allows for specific reward formulation to achieve desired response generation. This allowed $\mathcal{K}$-PERM to guide LLMs like GPT 3.5 to produce personalized results. Our approach can be generalized to other domains and LLMs using appropriate goal-oriented and personalized datasets.

\section{Limitations}
\begin{itemize}
    \item We only evaluate $\mathcal{K}$-PERM on the FoCus dataset. To the best of our knowledge, only one dataset incorporates persona, context, and queries together. However, we believe that with the growing interests and advancements, more such datasets will be constructed, and the solutions evolved and generalized.
    \item Our methodology did not thoroughly experiment with all the state-of-the-art LLMs such as Llama and Mistral \cite{jiang2023mistral}. However, as our methods are model-agnostic, expanding the work to these models should yield considerable results.
\end{itemize} 
\bibliography{references}

\appendix
\section*{Appendix}
\section{Dataset}
\label{dataset}
We utilize the publicly available FoCus dataset, which consists of passages describing landmarks \cite{jang2022call}. The dataset includes 13,484 dialogs for training and validation, with an average of 5.6 rounds per dialog and approximately 7,715 Wikipedia landmarks. The dialogs contain a total of 75,971 utterances, with an average length of 24.0 per utterance. We split the dataset into three sets: train (10,284 samples), validation (1,600 samples), and test (1,600 samples), comprising 57,928, 9,008, and 9,035 utterances, respectively. The training set includes 36,472 knowledge-based utterances and 21,456 utterances with both persona and knowledge, coming from 4,918 landmarks. The validation and test sets contain 5,664 and 5,707 knowledge-based utterances and 3,344 and 3,328 utterances with both persona and knowledge, respectively. The validation set covers 1,414 Wikipedia landmarks, while the test set covers 1,383 landmarks. The dataset references "ground persona" and "ground knowledge," which represent the ground truth persona and passage selected by crowd workers. Additionally, all the questions in the dataset were rewritten using T5-CANARD, a query-rewriting model \cite{qianfull} (Table \ref{tab:retriever_perf} shows why question rewriting was needed).

\section{Related Work}
PersonaChat and PersonaChat 2.0 are chit-chat conversation datasets used to train conversational agents~\cite{liu2022personalized, wuguiding}. They incorporate encoder-decoder models, reinforcement learning, few-shot learning, and hierarchical attention mechanisms to improve personalization by considering persona and conversational history~\cite{li2016deep,qiu2021learning,young2022fusing}. In previous personalization work, external knowledge was not a significant focus until Mazumder et al. introduced retrieval as a sub-task in PersonaChat to enhance personalization~\cite{majumder2021unsupervised}. However, their method was not specifically evaluated on retrieval performance, using ROC Stories as a proxy knowledge source. This approach did not adequately capture personalization, especially for goal-oriented or information-seeking dialogues. $\mathcal{K}$-PERM addresses this gap by introducing retrieval augmented generation as a practical baseline to improve personalization~\cite{gaur2022iseeq, lewis2020retrieval}. Unlike previous approaches that fine-tuned models on PersonaChat, $\mathcal{K}$-PERM uses base generative models for evaluation and achieves desired behaviors through reinforcement learning as the evaluator~\cite{lipton2018bbq, huang2023personalized}. Additionally, $\mathcal{K}$-PERM introduces unique persona selection, making it the first realistic response generation model for real-time goal-oriented or information-seeking dialogues.

\section{\textbf{Human Evaluation}} 
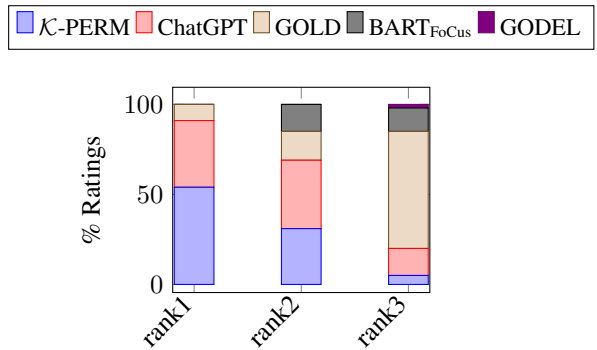
\begin{figure}
    \centering
    \begin{tikzpicture}
        \begin{axis}[
            ybar stacked,
            bar width=15pt,
            legend style={at={(0.5,1.4)},
            anchor=north,legend columns=-1},
            ylabel={\% Ratings},
            symbolic x coords={rank1, rank2, rank3},
            xtick=data,
            x tick label style={rotate=45, anchor=east},
        ]
        \addplot+[ybar] plot coordinates {(rank1,54) (rank2,31) (rank3,5)};
        \addplot+[ybar] plot coordinates {(rank1,37) (rank2,38) (rank3,15)};
        \addplot+[ybar] plot coordinates {(rank1,9) (rank2,16) (rank3,65)};
        \addplot+[ybar] plot coordinates {(rank1,0) (rank2,15) (rank3,13)};
        \addplot+[ybar] plot coordinates {(rank1,0) (rank2,0) (rank3,2)};
        \legend{\footnotesize{\strut $\mathcal{K}$-PERM}, \footnotesize{\strut ChatGPT}, \footnotesize{\strut GOLD}, \footnotesize{\strut $\text{BART}_\text{FoCus}$}, \footnotesize{\strut GODEL}}
    \end{axis}
    \end{tikzpicture}
    \caption{$\mathcal{K}$-PERM was preferred 32\% more than ChatGPT by the annotators based on a blind evaluation of 90 queries taken randomly from the FoCus dataset. GOLD is the ground truth in FoCus Dataset.}
    \label{fig:qual_eval}
\end{figure}

We conducted a blind evaluation of 90 responses generated from 5 systems. As the task was trivial, the authors instructed the annotators verbally. The task was made available on a participant pool management system which gives credits to students completing the task. We considered the five systems $\mathcal{K}$-PERM, $\text{BART}_\text{FoCus}$, ChatGPT, GODEL, and Ground Truth. Our model, $\mathcal{K}$-PERM, consistently stood out as a top performer by securing first place in 11 instances and maintaining a strong second position in 68 instances. This underscores its exceptional performance. The Top 3 results and percentage of times any specific model is selected are depicted in Figure \ref{fig:qual_eval}. Table \ref{tab:human} shows a sample from our human evaluation. \footnote{More examples are present \textcolor{blue}{\texttt{\small https://shorturl.at/uP023}}}. 

\begin{table*}[t]
  \centering
  \begin{tabular}{lcccccccc}
    \toprule
    \textbf{Models} & \textbf{NUBIA} & \textbf{SR} & \textbf{Contraction (\%)} & \textbf{Irrelevancy (\%)} & \textbf{LA (\%)} \\
    \midrule
    $BART_\text{FoCus}$ & 0.10 & 1.601 & 28.72 & 72.71 & 16.75 \\
    \midrule
    $\mathcal{K}$-PERM (GK only) & 0.24 & 2.20 & 11.78 & 74.4 & 13.5 \\
    $\mathcal{K}$-PERM (All P + $\mathcal{Z}_k$) & 0.22 & 1.90 & 13.18 & 70.1 & 17.5 \\
    $\mathcal{K}$-PERM (GP + $\mathcal{Z}_k$) & \textbf{0.34} & \textbf{2.80} & \textbf{11.87} & \textbf{66.38} & \textbf{21.75} \\
    $\mathcal{K}$-PERM (SP + $\mathcal{Z}_k$) & \underline{0.31} & \underline{2.42} & \underline{11.85} & \underline{69.61} & \underline{19.44} \\
    \bottomrule[1.5pt]
  \end{tabular}
  \caption{Performance of $\mathcal{K}$-PERM using NUBIA Score. Results are compared with baseline $\text{BART}_\text{FoCus}$. Bold-faced are best (GP), and Underlined are second-best ($P_\text{select}$). LA: Logical Agreement, SR: Semantic Relatedness.}
  \label{tab:ablation}
\end{table*}

\begin{table*}[!ht]
    \centering
    \begin{tabular}{l|p{12cm}}
        \toprule[1.5pt]
         Wikipedia Topic & /wiki/Mount\_Monadnock\\
         Query & Where is the mountain? \\
         \midrule
         \multicolumn{2}{c}{Personalized Response Generation: Example 1} \\ \midrule
         Ground Truth & Mount Monadnock is a mountain in the towns of Jaffrey and Dublin, New Hampshire. \\
         $\text{BART}_\text{FoCus}$ & Mount Monadnock is at New Hamsphire.\\
         GODEL & It's near the Rivers and the Masses. \\
         ChatGPT & Based on the context and your persona, you have already visited Mount Monadnock! It is located in the towns of Jaffrey and Dublin, New Hampshire. \\
         $\mathcal{K}$-PERM* & Mount Monadnock is a \textbf{3,165-ft (965 m)} mountain in the towns of Jaffrey and Dublin, New Hampshire, a place you have visited. \\ \midrule
         \multicolumn{2}{c}{Personalized
         Response Generation: Example 1} \\ \midrule
         Wikipedia Topic & /wiki/Assiniboine\_Park\_Zoo \\
         Query & Where is this place?\\
         ChatGPT & ``The query "Where is this place?'' is referring to the Assiniboine Park Zoo in Winnipeg, Manitoba, Canada. If you like Canada, you should definitely check out this zoo when you can! It's been a staple of the city since 1904! \\
         $\mathcal{K}$-PERM$\dagger$ & Assiniboine Park Zoo is an \textbf{80-acre (32 ha)} zoo in Manitoba, Canada that was established in 1904.\\
         \bottomrule[1.5pt]
    \end{tabular}
    \caption{Human Assessment: $\mathcal{K}$-PERM did better than ChatGPT in terms of personalized response generation. The personas ``I have been to New Hampshire'' (top) and ``I like visiting Canada'' (bottom) were used in the examples above. We bold face numeric values to show focus on the correctness in information delivery using a retrieval-augmented generation mechanism.}
    \label{tab:human}
\end{table*}

\section{Fine-Tuning MPNET for DPR \label{sub:mpnet-ft}}

\begin{table*}[!ht]
\centering
\begin{tabular}{lcccccccc}
\hline
\multirow{2}{*}{\textbf{Model}} & \multicolumn{4}{c}{\textbf{Original Query}} & \multicolumn{4}{c}{\textbf{Rewritten Query}} \\ \cmidrule{2-9}
\# Passages & 5 & 10 & 15 & 20 & 5 & 10 & 15 & 20 \\ \midrule[1.5pt]
TFIDF & 0.385 & 0.412 & 0.396 & 0.395 & 0.403 & 0.406 & 0.406 & 0.406 \\
(TFIDF + ptECE) & 0.422 & 0.434 & 0.430 & 0.427 & 0.441 & 0.452 & 0.457, & 0.445 \\
BM25 & 0.398 & 0.405 & 0.410 & 0.401 & 0.411 &  0.416 & 0.416 & 0.412 \\
(BM25 + ptECE) & 0.471 & 0.465 & 0.480 & 0.474 & 0.475 & 0.477 & 0.493 &0.467 \\
Mpnet & 0.542& 0.558 & 0.613 & 0.621& 0.540 & 0.560 & 0.608 & 0.604 \\
Mpnet\textsubscript{Fine tuned} & \textbf{0.642} & \textbf{0.658} & \textbf{0.683} & \textbf{0.661} &  \textbf{0.640} & \textbf{0.660} & \textbf{0.688} & \textbf{0.684} \\
ColBERT+X\cite{lawrie2022multilingual} & \underline{0.588} & \underline{0.584} &  \underline{0.601} &  \underline{0.601} & \underline{0.598} & \underline{0.602} &  \underline{0.605} & \underline{0.601} \\
ColBERT \cite{khattab2020colbert} & 0.560 & 0.591 &  0.603 & 0.60 &  0.572 & 0.602 & 0.602 & 0.598 \\
\bottomrule[1.5pt]
\end{tabular}
\caption{Evaluating knowledge retrievers over the indexed landmark Wikipedia articles in the FoCus dataset. Each score represents the maximum BERTScore computed between the retrieved passages and ground truth knowledge in FoCUS. ptECE: pre-trained ELECTRA Cross Encoder \cite{clark2020electra}. Bold: Best, Underlined: 2nd best.}
\label{tab:retriever_perf}
\end{table*}

We used contrastive fine-tuning as described in \cite{song2020mpnet}. Next, we employ a combination of locality-sensitive hashing (LSH) and Facebook AI Semantic Search (FAISS), which uses \textbf{Maximum Inner Product Search (MIPS)} \cite{johnson2019billion} \textbf{to efficiently obtain} $\mathbf{\mathcal{Z}^{U^t}_K}$. We evaluated the knowledge retriever's efficiency using BERTScore, which is commonly used to compare retriever-augmented generations\cite{lim2023you}.

We varied $K$ between 5 and 20 and observed high similarity (BERTScore) with ground truth passages (from the FoCus dataset) at $K=10$. Therefore, we agreed that the appropriate number of retrieved passages would be 10. We also experimented with different Sentence Transformer models for dense encodings other than Mpnet and standard retrievers such as TF-IDF and BM25 (refer to Appendix Table \ref{tab:retriever_perf}).

\section{Training Process}
\label{training}
We used BART as our auto-regressive encoder-decoder model for personalized response generation conditioned upon the query and retrieved knowledge. The BART model was fine-tuned following the procedure of training BART for conversation summarization but with special tags: \textit{<question>}, \textit{<knowledge>}, \textit{<history>}, and \textit{<persona>}, appropriately placed for response generation \cite{chen2021structure}. Fine-tuning of BART was performed on the entire training data, and validation was set for four epochs, taking approximately 16 hours with 1 NVIDIA T4 GPU. For generating a response, we used beam search with beam size 5. We selected beam search over nucleus sampling for stability in a generation \cite{shaham2022cross}.

\end{document}